\documentclass[letter, twocolumn, superscriptaddress, footinbib, aps, prd, 10pt]{revtex4-2}
\pdfoutput=1

\usepackage{lmodern}

\usepackage[normalem]{ulem}
\usepackage[T1]{fontenc}
\usepackage[latin9]{inputenc}
\usepackage[verbose,tmargin=0.75in,bmargin=1in,lmargin=0.75in,rmargin=0.75in,columnsep=0.25in]{geometry}
\usepackage{subfigure, adjustbox, bm, xcolor}
\usepackage[noEucal]{main}
\usepackage{aas_macros}

\def\edge{{\text{edge}}}

\begin{document}

\preprint{}

\hfill CALT-TH 2024-018

\title{Diamond of Infrared Equivalences in Abelian Gauge Theories}

\author{Temple He}
\affiliation{Walter Burke Institute for Theoretical Physics \\ California Institute of Technology, Pasadena, CA 91125 USA}
\author{Prahar Mitra}
\affiliation{Institute for Theoretical Physics, University of Amsterdam \\
Science Park 904, Postbus 94485, 1090 GL Amsterdam, The Netherlands}
\author{Kathryn M. Zurek}
\affiliation{Walter Burke Institute for Theoretical Physics \\ California Institute of Technology, Pasadena, CA 91125 USA}

\date{\today}

\begin{abstract}
We demonstrate a tree-level equivalence between four distinct infrared objects in $(d+2)$-dimensional abelian gauge theories. These are ($i$) the large gauge charge $Q_\ve$ where the function $\ve$ on the sphere parameterizing large gauge transformations is identified with the Goldstone mode $\t$ of spontaneously broken large gauge symmetry; ($ii$) the soft effective action that captures the dynamics of the soft and Goldstone modes; ($iii$) the edge mode action with Neumann boundary conditions; and ($iv$) the Wilson line dressing of a scattering amplitude, including a novel dressing for soft photons, which have local charge distributions despite having vanishing global charge. The promotion of the large gauge parameter to the dynamical Goldstone and the novel dressing of soft gauge particles give rise to intriguing possibilities for the future study of infrared dynamics of gauge theories and gravity.
\end{abstract}

\maketitle

\section{Introduction}

Although soft theorems were discovered more than half a century ago \cite{Low:1958sn, Weinberg:1965aa, Berends:1988zn}, their full potential was not appreciated until relatively recently, when they were shown to be Ward identities associated with asymptotic symmetries \cite{Strominger:2013lka, Strominger:2013jfa, He:2014laa, He:2014cra, Lysov:2014csa, He:2015zea, Campiglia:2015qka, Kapec:2015ena, Campiglia:2015kxa, Campiglia:2016hvg, He:2020ifr}. Advancement in our study of asymptotic symmetries in asymptotically flat spacetimes has led to a deeper understanding of the infrared (IR) sector of quantum field theories (QFTs), celestial holography, and new memory effects (see \cite{Strominger:2017zoo, Raclariu:2021zjz, Pasterski:2021rjz, Pasterski:2021raf, McLoughlin:2022ljp} for a review and a complete list of references).

It has long been known that in theories with massless particles, scattering amplitudes admit a soft-hard factorization \cite{Bassetto:1983mvz, Feige:2014wja}. More precisely, a scattering amplitude with $m$ soft (low energy) particles and $n$ hard particles (high energy) factorizes as \footnote{Strictly speaking, this factorization has been proven to hold for abelian gauge theories without massless charged particles.}
\begin{align}
\label{soft_factorization}
\CA_{m+n} = e^{-\G} \SS_m {\tilde \CA}_n,
\end{align}
where $\SS_m$ is the contribution of real soft particles, $\G$ the contribution of virtual soft particles, and ${\tilde \CA}_n$ the amplitude involving only hard states (real and virtual). IR divergences in four-dimensional theories with massless particles are captured by the factor $e^{-\G}$, which is formally zero. The precise form of the soft factorization Eq.~\eqref{soft_factorization} is typically determined by explicitly evaluating the associated Feynman diagrams, which involves a rather convoluted calculation. However, given the infinite-dimensional asymptotic symmetries that constrain the IR sector of a QFT, one can hope that the soft factor $e^{-\G} \SS_m$ can be obtained more directly from symmetry principles. In \cite{Kapec:2021eug}, it was shown that this is the case for abelian gauge theories.

In these theories, the asymptotic symmetries are gauge transformations that act non-trivially at infinity, which we refer to as \emph{large gauge transformations} (LGTs). Unlike gauge transformations that act trivially at infinity (which are redundancies of the theory), LGTs are ``real'', and physical states are generally not annihilated by the corresponding Noether charge $Q_\ve$, where $\ve$ is the gauge parameter. Indeed, such a non-trivial charge implies that gauge theories have an infinite degeneracy of vacua, all of which are related by $Q_\ve$. Moreover, the charge is conserved, and its Ward identity is precisely Weinberg's leading soft photon theorem \cite{He:2014cra, Campiglia:2015qka, Kapec:2015ena, Kapec:2014zla, He:2019jjk}. By exploiting these infinite-dimensional large gauge symmetries, the authors of \cite{Kapec:2021eug} showed that the soft factor $e^{-\G} \SS_m$ in $(d+2)$-dimensional abelian gauge theories is exactly reproduced by a $d$-dimensional action (which we refer to as the \emph{soft effective action}) that describes the effective dynamics of the soft modes and their interaction with the hard modes.

In this paper, we uncover a tree-level equivalence among four distinct objects that appear in the study of IR physics in abelian gauge theories. These are depicted as the four corners of what we refer to as the \emph{infrared diamond} in Fig.~\ref{fig:diamond}. The top corner is the aforementioned soft effective action \footnote{The full soft effective action appearing in \cite{Kapec:2021eug} contains another term $\a \int \frac{\dt^d x}{(2\pi)^d} [ N_a(x) ]^2$, which arises from one-loop effects. In this paper, we restrict our discussion to tree-level and consequently ignore this term.\label{footnote:loop-sea}}. The left corner is the large gauge charge with the gauge parameter $\ve$ elevated to the Goldstone operator $\t$. The right corner is the edge mode action \footnote{Dynamics of edge modes in gauge theory was first discussed in \cite{Donnelly:2011hn, Donnelly:2014gva, Donnelly:2014fua, Donnelly:2015hxa}. The relationship between the edge mode action and the soft effective action has been discussed recently in \cite{Chen:2023tvj, Chen:2024kuq, Ball:2024hqe, He:2024ddb}.}, which is the soft limit of the gauge theory action (with appropriate boundary terms) evaluated on-shell. Finally, the bottom corner is a soft Wilson line dressing. This includes a dressing factor for both the hard charged particles, previously discussed in \cite{Laenen:2008gt, White:2011yy, Feige:2014wja, Nguyen:2023ibj, Bonocore:2021qxh}, and (crucially) for soft photons, which carry local, but not global, electric charge.

\begin{figure*}[ht!] 
\centering
\label{fig:diamond}
\begin{tikzpicture}
\draw[line width = 2pt, color = blue] (0,0) -- (2,2) -- (0,4) -- (-2,2) -- cycle;
\draw[line width = 2pt, color = blue, dotted] (0,0) -- (0,4);
\node[circle, draw=blue, fill=blue, inner sep=2pt] at (0,0) {};
\node[circle, draw=blue, fill=blue, inner sep=2pt] at (2,2) {};
\node[circle, draw=blue, fill=blue, inner sep=2pt] at (0,4) {};
\node[circle, draw=blue, fill=blue, inner sep=2pt] at (-2,2) {};
\node[fill=white] (n1) at (0,2) {Sec.~\ref{ssec:wilson=soft}};
\node[shift={(45:0.01cm)},anchor=south west] (n2) at (1,3) {Sec.~\ref{ssec:edge=soft}};
\node [shift={(135:0.01cm)}, anchor=south east] (n3) at (-1,3) {Sec.~\ref{ssec:sea=charge}};
\node[black] at (0,4.5) {$S^\text{tree}_{\text{eff}} = \displaystyle{ - \frac{i}{2c_{1,1}}  \oint_{\mss^d} } \!\! \dt^d x \, {\wt C}^a(x) [ N_a(x) - \CJ_a(x) ]$};
\node[black, right=0.25cm] at (2, 2.5) {$S_{\text{edge}} = \bigg[  - \displaystyle\int_\CM \dt^{d+2} X \sqrt{-g}  \left( \frac{1}{4e^2} F^{\mu\nu} F_{\mu\nu} + A^\mu J_\mu \right)$};
\node[black, right=0.25cm] at (3.5, 1.5)
{$+ \displaystyle{\frac{1}{e^2}   \int_{\S^+ - \S^-}}  \dt \S^\mu A^\nu F_{\mu\nu} \bigg] \bigg|_{\text{on-shell}}$};
\node[black, below=0.25cm] at (0,0) {$\CW_S^\tree \CW_M^\tree = \exp\left[ i \displaystyle\oint_{\mathbb S^d} \!\! \dt^dx \,\rho_{\text{total}}(x)  \t(x) \right]  $};
\node[black, left=0.25cm] at (-2.0,2) {$Q^+_{\t} - Q^-_{\t} = \displaystyle - \frac{1}{2e^2} \oint_{\CI^+_- + \CI^-_+} \!\! \dt S^{\mu\nu} \, \t F_{\mu\nu}$};
\end{tikzpicture}
\caption{The IR diamond above captures the tree-level equivalence among four distinct IR objects.  These objects are, beginning on the left corner of the diamond and moving clockwise, ($i$) the large gauge charge $Q^\pm_\ve$ with $\ve = \t$ (Eq.~\eqref{charge}) so that it is parametrized by the dynamical Goldstone field $\theta$; ($ii$) the soft effective action (Eq.~\eqref{sea}) which captures the dynamics of the soft photon $N_a$ and the Goldstone mode $C_a = \p_a \t$; ($iii$) the (soft limit of the) edge mode action (Eq.~\eqref{action_1}) previously analyzed in \cite{He:2024ddb}; and ($iv$) a product of tree-level Wilson line dressing of matter $\CW_M^\tree$ and a novel soft photon Wilson line dressing ${\cal W}_S^\tree$ (Eq.~\eqref{wilson-soft} or Eq.~\eqref{soft-dressing-final}). The label on each edge denotes the section where we prove the equivalence between the two endpoints.}
\end{figure*}

The rest of this paper is organized as follows. In Section~\ref{sec:prelims}, we review each of the corners of the diamond. In Section~\ref{sec:equivalence}, we prove the equivalence of the soft effective action to the other three corners of the diamond. Finally, in Section~\ref{sec:discussion}, we comment on the implications of these results and future directions.

\section{Preliminaries}
\label{sec:prelims}

\paragraph{Coordinate Conventions:} In this paper, we study $U(1)$ gauge theories in $D=d+2$ dimensional Minkowski spacetime $\CM$. We work in flat null coordinates $X^\mu = ( u , x^a , r)$, related to Cartesian coordinates $X^A$ via $X^A(u,x,r) = r {\hat q}^A(x) + u n^A$, where
\begin{equation}
\begin{split}
\label{mompar}
{\hat q}^A(x) &= \left( \frac{1+x^2}{2} , x^a , \frac{1-x^2}{2} \right)  \\
n^A &= \left( \frac{1}{2}, 0^a ,-\frac{1}{2} \right) . 
\end{split}
\end{equation}
In flat null coordinates, the metric of Minkowski spacetime takes the form
\begin{align}
\dt s^2 = - \dt u \, \dt r + r^2 \d_{ab} \, \dt x^a \,\dt x^b.
\end{align}
The asymptotic null boundaries $\CI^\pm$ of Minkowski spacetime are located at $r \to \pm \infty$. This is coordinatized by $(u,x^a)$ and has the topology of $\mrr \times \mss^d$. The past (future) boundary of $\CI^+$ ($\CI^-$) is located at $u \to -\infty$ ($u \to + \infty$) and is denoted by $\CI^+_-$ ($\CI^-_+$). The future and past timelike boundaries of $\CM$ are denoted by $i^\pm$. The asymptotic past and future Cauchy slices of the spacetime are $\S^\pm = \CI^\pm \cup i^\pm$, and their corresponding boundaries are $\p\S^\pm = \CI^\pm_\mp$. In these coordinates, the relevant future-pointing area elements are given by
\begin{equation}
\begin{split}
\label{area_elements}
\dt \S^\mu |_{\CI^\pm} &= - \d^\mu_u |r|^d \, \dt u \, \dt^d x \\
\dt S^{\mu\nu} |_{\CI^\pm_\mp} &= 4 \d^{[\mu}_u \d^{\nu]}_r  |r|^d \, \dt^d x  . 
\end{split}
\end{equation}

\paragraph{Scattering Amplitudes:} We denote by $\CO_k$ the operator that inserts the $k$th external state in a scattering amplitude. For conciseness, we suppress all of its arguments, which include the momentum $p_k^A$ (satisfying $p_k^2 = - m_k^2$), electric charge $Q_k \in \mzz$, Lorentz spin indices, flavor indices, etc. An $n$-point scattering amplitude evaluated in the standard Lorentz-invariant QFT vacuum can then be written as \footnote{Gauge theories have infinitely many vacuum states, and it is possible to consider scattering amplitudes with different $in$ and $out$ vacua. We do not explore this possibility here}
\begin{equation}
\begin{split}
\label{amplitude}
\CA_n = \avg{ \CO_1 \cdots \CO_n } .
\end{split}
\end{equation}
We define the \emph{soft photon operator} as
\begin{equation}
\begin{split}
\label{Na_def}
N_a(x) &\equiv \frac{1}{2e} \left( \lim_{\o \to 0^+} + \lim_{\o \to 0^-} \right) \left[ \o \CO_a(\o {\hat q}(x)) \right] ,
\end{split}
\end{equation}
where $\CO_a(\o {\hat q}(x))$ inserts a photon with momentum $\o {\hat q}^A(x)$ and polarization $\ve_a^A(x) = \p_a {\hat q}^A(x)$, and $e$ is the gauge coupling constant.

\paragraph{Soft Factorization:} Insertions of the soft photon operator Eq.~\eqref{Na_def} are universal and are described by Weinberg's leading soft photon theorem \cite{Weinberg:1965aa}
\begin{equation}
\begin{split}
\label{soft_ph_thm}
\avg{ N_a(x) \CO_1 \cdots \CO_n } = \CJ_a(x) \avg{ \CO_1 \cdots \CO_n } ,
\end{split}
\end{equation}
where
\begin{equation}
\begin{split}
\label{Ja_def}
\CJ_a(x) \equiv \p_a \sum_{k=1}^n Q_k \ln | p_k \cdot {\hat q}(x) | .
\end{split}
\end{equation}
Here, we are working in the standard convention where outgoing particles have charge $Q_k$ while incoming particles have charge $-Q_k$.

\paragraph{Large Gauge Symmetries:} 

Abelian gauge theories are invariant under gauge transformations that act on the gauge field $A$ via \footnote{We normalize the gauge field so that the pure Maxwell action is $S_\text{Maxwell} = - \frac{1}{2e^2} \int_\CM F \wedge \star F$.}
\begin{equation}
\begin{split}\label{LGT}
A_\mu(X) \to A_\mu(X) + \nabla_\mu  \ve(X) , \quad \ve(X) \sim \ve(X) + 2\pi .
\end{split}
\end{equation}
When $\ve|_{\p \S^\pm}=0$, Eq.~\eqref{LGT} has no physical effect and is the usual small gauge redundancy. When $\ve|_{\p \S^\pm} \neq 0$, the transformations Eq.~\eqref{LGT} are the physical large gauge transformations. The Noether charge that generates these transformations is \cite{He:2014cra} 
\begin{equation}
\begin{split}
\label{charge}
    Q^\pm_\ve &= \mp \frac{1}{e^2} \oint_{\CI^\pm_\mp} \!\! \ve \star F = \mp \frac{2}{e^2} \oint_{\CI^\pm_\mp} \!\! \dt^d x \, |r|^d \ve F_{ur} ,
\end{split}
\end{equation}
where $F_{\mu\nu} = \p_\mu A_\nu - \p_\nu A_\mu$. The Ward identity associated with these large gauge symmetries is the leading soft photon theorem Eq.~\eqref{soft_ph_thm} \cite{He:2014cra, Campiglia:2015qka, Kapec:2014zla, Kapec:2015ena, He:2019jjk}. To show this, we use  Maxwell's equation, namely
\begin{equation}
\begin{split}
\label{Maxwell_Eq}
    \dt \star F = (-1)^d e^2 \star J \quad\Longleftrightarrow\quad  \nabla^\mu F_{\mu\nu} = e^2 J_\nu ,
\end{split}
\end{equation}
to rewrite the charge Eq.~\eqref{charge} as
\begin{equation}
\begin{split}
\label{charge_decomposition}
    Q^\pm_\ve &= \frac{1}{e^2} \int_{\S^\pm} \dt(\ve \star F) \\
    &= \frac{1}{e^2} \int_{\S^\pm} \dt \ve \wedge \star F + (-1)^d \int_{\S^\pm} \ve \star J. 
\end{split}
\end{equation}
We denote the first term by $Q_\ve^{\pm S}$ and the last term by $Q_\ve^{\pm H}$, so that
\begin{align}\label{eq:soft-and-hard}
\begin{split}
    Q_\ve^{S \pm} &= \frac{1}{e^2} \int_{\S^\pm} \dt \ve \wedge \star F = \frac{1}{e^2} \int_{\S^\pm} \dt \S^\mu \nabla^\nu \ve F_{\mu\nu} \\
    Q_\ve^{H \pm} &= (-1)^d \int_{\S^\pm} \ve \star J = - \int_{\S^\pm} \dt\S^\mu \, \ve J_\mu.
\end{split}
\end{align}
The Ward identity $Q_\ve^+ = Q_\ve^-$ then implies
\begin{equation}
\begin{split}
\label{ward_id}
&\avg{ ( Q_\ve^{+S} - Q_\ve^{-S} )  \CO_1 \cdots \CO_n } \\
&\qquad \qquad \qquad = - \avg{ ( Q_\ve^{+H} - Q_\ve^{-H} )  \CO_1 \cdots \CO_n } . 
\end{split}
\end{equation}
It was shown in \cite{He:2014cra, He:2019jjk, He:2023bvv} that the soft charge $Q_\ve^{\pm S}$ is the soft photon operator Eq.~\eqref{Na_def} smeared over $\mss^d$. With a clever choice of the gauge parameter $\ve$ (see \cite{He:2019jjk} for details), the left-hand-side of Eq.~\eqref{ward_id} can be made identical to that of Eq.~\eqref{soft_ph_thm}. On the other hand, the hard charges are quadratic in the charged matter fields, so they act on the hard states and transform them. It was shown in \cite{He:2014cra, Campiglia:2015qka, Kapec:2014zla, Kapec:2015ena, He:2019jjk} that this precisely reproduces the right-hand-side of Eq.~\eqref{soft_ph_thm}.

\paragraph{Soft Effective Action:} 

In \cite{Kapec:2021eug}, Kapec and Mitra identified the relevant soft degrees of freedom in abelian gauge theories and, using the large gauge symmetry of the system, constructed an effective action for these modes. At tree-level (see Footnote \cite{Note1}), this so-called \emph{soft effective action} is given by 
\begin{equation}
\begin{split}
\label{sea}
S^\text{tree}_{\text{eff}} &= - \frac{i}{2c_{1,1}} \oint_{\mathbb S^d} \!\! \dt^d x \, {\wt C}^a(x) [ N_a(x) - \CJ_a(x) ] ,
\end{split}
\end{equation}
where $N_a(x)$ and $\CJ_a(x)$ are respectively defined in Eq.~\eqref{Na_def} and Eq.~\eqref{Ja_def}, and $C_a(x)$ is the boundary value of the gauge field, i.e., $C_a(x) \equiv A_a |_{\CI^+_-} (x) = A_a |_{\CI^-_+} (x) $ \footnote{The second equality holds due to the matching condition given in \cite{He:2014cra}.}. In the absence of magnetic charges, we have
\begin{equation}
\begin{split}
\label{Ca_def}
C_a(x) = \p_a \t(x)  , \qquad \t(x) \sim \t (x) + 2 \pi  . 
\end{split}
\end{equation}
Now, in the perturbative Lorentz invariant vacuum, we have $\avg{\t(x)}=0$ \cite{He:2014cra,He:2020ifr}. However, under an LGT given in Eq.~\eqref{LGT}, $\t(x) \to \t(x) + \ve(x)$ for $\ve(x) \equiv \ve|_{\p\S^\pm}(x)$, implying that the Lorentz invariant vacuum state spontaneously breaks the large gauge symmetry \emph{and} that $\t(x)$ is the corresponding Goldstone mode.

Finally, the tilde on $C_a$ denotes the shadow transform. For a vector field of dimension $\D=1$, this is defined to be
\begin{equation}
\begin{split}
{\wt C}_a (x) &\equiv \int \dt^d y \frac{\d_{ab} - 2 \frac{(x-y)_a (x-y)_b }{(x-y)^2} }{ [ ( x - y )^2 ]^{d-1} } C^b(y)  . 
\end{split}
\end{equation}
Up to normalization, the shadow transform is its own inverse:
\begin{equation}
\begin{split}
\wt{\wt V}_a(x) &= c_{1,1} V_a(x) .
\end{split}
\end{equation}
The constant $c_{1,1}$ appears in the action Eq.~\eqref{sea}, and it was shown in \cite{Kapec:2021eug} that the action Eq.~\eqref{sea} correctly reproduces the soft theorem Eq.~\eqref{soft_ph_thm}.

\section{Equivalence Relations}
\label{sec:equivalence}

In the following three subsections, we will demonstrate the equivalences of the IR diamond shown in Fig.~\ref{fig:diamond}.

\subsection{Large Gauge Charge = Soft Effective Action }\label{ssec:sea=charge}

We first prove the upper left edge in Fig.~\ref{fig:diamond} by demonstrating
\begin{equation}
\begin{split}
\label{equivalence_1}
Q^+_\t - Q^-_\t  =  i S^\text{tree}_\text{eff} . 
\end{split}
\end{equation}
To this end, we begin by using Eq.~\eqref{charge} to write
\begin{equation}\label{Q-dif}
\begin{split}
& Q^+_\t - Q^-_\t \\
&\quad = - \frac{2}{e^2} \oint_{\mss^d}\!\! \dt^d x\, \t \left[ ( |r|^d F_{ur} ) |_{\CI^+_-} + ( |r|^d F_{ur} ) |_{\CI^-_+} \right] .
\end{split}
\end{equation}
Following \cite{He:2019jjk}, we decompose the field strength into a radiative part (that solves free Maxwell's equations) and a Coulombic part (that, via a Green's function, describes the interaction of the gauge field with charged matter fields), so that
\begin{equation}
\begin{split}
( |r|^d F_{ur} ) |_{\CI^\pm_\mp} = ( |r|^d F_{ur}^{R\pm} ) |_{\CI^\pm_\mp} + ( |r|^d F_{ur}^{C\pm} ) |_{\CI^\pm_\mp} .
\end{split}
\end{equation}
The $\pm$ superscript distinguishes the Green's function that is used to evaluate the Coulombic part of the field: $+$ for the advanced Green's function and $-$ for the retarded one.

It was shown in \cite{He:2023bvv} that the radiative part of the field strength can be written in terms of the soft photon operator Eq.~\eqref{Na_def} as \footnote{$N_a$ as defined in this paper is $N_a^+ - N_a^-$ in \cite{He:2023bvv}.}
\begin{equation}
\begin{split}
\label{Radiative_REF}
( |r|^d F^{R+}_{ur} ) |_{\CI^+_-} + ( |r|^d F^{(R-)}_{ur} ) |_{\CI^-_+} &=  \frac{e^2}{4 c_{1,1}} \p^a {\wt N}_a .
\end{split}
\end{equation}
Additionally, \cite{He:2019jjk} showed that the Coulombic part of the field strength satisfies the equation
\begin{equation}
\begin{split}
\label{Coulombic_REF}
&( |r|^d F_{ur}^{C+} ) |_{\CI^+_-} + ( |r|^d F_{ur}^{C-} ) |_{\CI^-_+}  \\
&\qquad = - \frac{e^2}{2} \int_\mrr  \dt u  \left[ ( |r|^d J_u ) |_{\CI^+}  - ( |r|^d J_u ) |_{\CI^-}  \right] \\
&\qquad \qquad \qquad + \left[ ( |r|^d F_{ur}^{C+} ) |_{\CI^+_+} + ( |r|^d F_{ur}^{C-} ) |_{\CI^-_-}  \right] . 
\end{split}
\end{equation}
Utilizing an identify proven in \cite{He:2024ddb}, the above expression is related to the soft factor given in Eq.~\eqref{Ja_def} via \footnote{The appearance of the shadow transform ensures the scaling dimensions on both sides of Eq.~ \eqref{Coulombic_REF2} to match while preserving Lorentz symmetry \cite{Kapec:2017gsg}.}
\begin{equation}\label{Coulombic_REF2}
\begin{split}
( |r|^d F_{ur}^{C+} ) |_{\CI^+_-} + ( |r|^d F_{ur}^{C-} ) |_{\CI^-_+}  = - \frac{e^2}{4c_{1,1}} \p^a {\wt \CJ}_a(x) . 
\end{split}
\end{equation}
Substituting Eq.~\eqref{Radiative_REF} and Eq.~\eqref{Coulombic_REF2} into Eq.~\eqref{Q-dif}, we obtain
\begin{equation}
\begin{split}
\label{charge_LHS}
Q_\t^+ - Q_\t^- &= - \frac{1}{2c_{1,1}} \oint_{\mss^d}\!\! \dt^d x\, \t(x) \p^a [ {\wt N}_a(x) - {\wt \CJ}_a(x) ]  \\
&= \frac{1}{2c_{1,1}} \oint_{\mss^d}\!\! \dt^d x\, {\wt C}^a (x)  [ N_a(x) - \CJ_a(x) ] ,
\end{split}
\end{equation}
where in the second equality, we first integrated by parts and then used the shadow transform property
\begin{equation}
\begin{split}
\oint_{\mss^d}\!\! \dt^d x\, C^a(x) {\wt C}'_a(x) = \oint_{\mss^d}\!\! \dt^d x\, {\wt C}^a(x) C'_a(x) . 
\end{split}
\end{equation}
Recalling Eq.~\eqref{sea}, we see from Eq.~\eqref{charge_LHS} that indeed Eq.~\eqref{equivalence_1} holds.

\subsection{Soft Effective Action = Edge Mode Action}
\label{ssec:edge=soft}

Next, we consider the upper right edge of the diamond in Fig.~\ref{fig:diamond}.  Donnelly and Wall showed in \cite{Donnelly:2014fua} that in a gauge theory, the entanglement entropy receives an extra contribution from the edge modes that live on the codimension-two entangling surface. The dynamics of these edge modes is described by a so-called edge mode action, which is the Maxwell action with appropriate boundary terms evaluated on-shell. An analogous edge mode action was considered in \cite{He:2024ddb}, albeit with different boundary conditions compared to \cite{Donnelly:2014fua}, and is given by
\begin{equation}
\begin{split}
\label{action_1}
    S_\edge &= - \int_\CM \dt^{d+2} X \sqrt{-g}  \left[ \frac{1}{4e^2} F^{\mu\nu} F_{\mu\nu} + A^\mu J_\mu \right] \\
    &\qquad  + \frac{1}{e^2} \int_{\S^+ - \S^-} \dt \S^\mu A^\nu F_{\mu\nu} .
\end{split}
\end{equation}
The boundary terms are required to ensure that the variational principle with radiative boundary conditions ($n^\mu \d F_{\mu\nu} |_{\S^\pm} = 0$) is well-defined, and $J_\mu$ is the conserved background matter charge current determined by integrating out the matter fields in the theory. It was shown in \cite{He:2024ddb} that the on-shell and soft limit of the above action precisely reproduces the soft effective action, which we shall now review. 

The edge modes whose action we are trying to construct are the soft photon mode Eq.~\eqref{Na_def} and the Goldstone mode Eq.~\eqref{Ca_def}. To ensure they appear in the action, it is necessary to decompose the gauge field as
\begin{equation}
\begin{split}
\label{gaugefield_decomp}
    A_\mu(X) = {\hat A}_\mu(X) + \nabla_\mu \t(X) ,
\end{split}
\end{equation}
where we have separated the pure gauge mode $\t(X)$ of the gauge field so that ${\hat A}_\mu(X)$ is large gauge invariant \footnote{This decomposition is uniquely specified by imposing a gauge-fixing condition on $A_\mu(X)$.}. Substituting this decomposition into Eq.~\eqref{action_1} and by taking the on-shell and soft limit, it was determined in \cite{He:2024ddb} that at tree-level,
\begin{align}\label{equiv2}
    S_\edge ~ \xrightarrow{\text{on-shell + soft}} ~ iS_\eff^\tree .
\end{align}
Indeed, it was shown in \cite{He:2024ddb} that this equivalence extends even to loop-level, but for our current purposes, Eq.~\eqref{equiv2} is the desired equivalence corresponding to the upper right edge in Fig.~\ref{fig:diamond}.

\subsection{Soft Effective Action = Wilson Line Dressing}
\label{ssec:wilson=soft}

The final equivalence we prove is the relationship between the soft effective action and a Wilson line dressing at tree-level, namely 
\begin{equation}
\begin{split}
\label{equivalence_3}
\exp [ - S^\text{tree}_{\text{eff}} ] = \CW^\text{tree}_S \CW^\text{tree}_M, 
\end{split}
\end{equation}
where $\CW_S$ is a dressing we introduce below for soft photons, $\CW_M$ the usual Wilson line dressing for charged matter particles, and the ``tree'' superscript indicates we are only evaluating their tree-level and soft contributions. There are two ways to interpret the Wilson line dressing above. The first interpretation is the conventional one, where the Wilson lines extend from a bulk point (say, the origin) to the celestial sphere on $\CI^\pm$. The second one is to view the Wilson lines as living wholly on the celestial sphere. We will see the equivalence Eq.~\eqref{equivalence_3} holds for both interpretations.

We begin with the first interpretation. In \cite{Feige:2014wja}, it was shown that the matter operator insertions $\CO_k$ appearing in the scattering amplitude are dressed with Wilson lines as $\CO_k = U_k \hat \CO_k$, with $\hat \CO_k$ being the ``bare'' (undressed) operators, and 
\begin{equation}
\begin{split}
\label{soft_dressing}
    U_k = \exp \bigg[ i Q_k \int_{\g_k} \dt X^\mu A_\mu(X) \bigg] ,
\end{split}
\end{equation}
where $\g_k$ is the worldline of the $k$th particle beginning from the origin \footnote{In the soft limit, the location of the point from which the Wilson line begins is inconsequential.}. The full Wilson line dressing of the scattering amplitude is given by $U_1 \cdots U_n$. The quantity that appears in Eq.~\eqref{equivalence_3} is the soft contribution to this Wilson line, namely 
\begin{equation}
\begin{split}
    \CW_M = U_1 \cdots U_n \big|_\text{soft} .
\end{split}
\end{equation}
To evaluate this, we rewrite the line integral in Eq.~\eqref{soft_dressing} in a more suggestive way. We parameterize the worldline $\g_k$ as $X_k^A(X^0) = ( X^0 , \vec{X}_k(X^0) )$ and write
\begin{equation}
\begin{split}
    U_k &= \exp\bigg[ i \eta_k Q_k  \int_0^\infty \dt X^0 \frac{\dt X_k^\mu}{\dt X^0}  A_\mu \left( X_k ( \eta_k X^0 ) \right)  \bigg] \\
    &= \exp\bigg[ -i \int_\CM \dt^{d+2} X \sqrt{-g} A_\mu(X) J^\mu_k(X) \bigg] ,
\end{split}
\end{equation}
where $\eta_k = +1$ ($-1$) if the $k$th particle is outgoing (incoming), and
\begin{equation}\label{point-current}
\begin{split}
J_k^\mu(X) = - \eta_k Q_k \t(\eta_k X^0) \frac{\dt X_k^\mu}{\dt X^0}  \d^{(d+1)} \left( \vec{X} - \vec{X}_k ( X^0 ) \right) 
\end{split}
\end{equation}
is the current corresponding to a charged point particle with $\t(X^0)$ being the Heaviside step function. Including the contribution from all the particles, we find
\begin{equation}\label{wilson-hard}
\begin{split}
\CW_M = \exp \bigg[ - i \int_\CM \dt^{d+2} X \sqrt{-g} A_\mu J^\mu \bigg] \bigg|_{\soft} ,
\end{split}
\end{equation}
where $J^\mu = \sum_{k=1}^n J_{k}^{\mu}$ is the \emph{total} conserved charged matter current. To further simplify this, we use the decomposition Eq.~\eqref{gaugefield_decomp}. As in \cite{He:2024ddb}, the $\hat A_\mu J^\mu$ term in the soft limit only contributes to the quantum piece of the soft effective action. Hence, only the $\nabla_\mu\t J^\mu$ term contributes at tree-level, and we have
\begin{align}\label{wilson-hard-1}
\begin{split}
    \CW_M^\tree &= \exp\bigg[ -i \int_{\CM} \dt^{d+2}X \sqrt{-g} \nabla_\mu \t J^\mu \bigg] \\
    &= \exp\bigg[ -i \int_{\S^+ - \S^-} \dt \S^\mu \, \t J_\mu \bigg] ,
\end{split}
\end{align}
where we used integration by parts and the current conservation equation $\nabla^\mu J_\mu = 0$, as well as the fact the matter current vanishes at spatial infinity $i^0$.

We next turn to the soft photon Wilson line dressing in Eq.~\eqref{equivalence_3}. At first glance, it seems that soft photons should not be dressed with any Wilson lines as they are uncharged, i.e., $Q_\text{photon} = 0$. However, as noted in \cite{He:2014cra}, while soft photons are globally neutral, they have a local charge distribution, which plays a crucial role in enforcing the local charge conservation law implied by the large gauge symmetry. It follows that soft photons \emph{must} be dressed not with a single Wilson line but with a ``Wilson line distribution.'' 

Indeed, analogous to Eq.~\eqref{point-current}, we define a conserved ``soft'' current
\begin{align}
\label{soft-current}
J_\mu^S = - \frac{1}{e^2} (\star \dt \star F )_\mu = \frac{1}{e^2} \n^\nu F_{\mu\nu} .
\end{align}
We note the current is trivially conserved since $F_{\mu\nu}$ is antisymmetric. The soft limit of the soft photon dressing is then just Eq.~\eqref{wilson-hard-1} with $J_\mu$ replaced with $J^S_\mu$, so that
\begin{align}\label{wilson-soft0}
\begin{split}
\CW^\text{tree}_S &= \exp \bigg[ - i\int_{\CM} \dt^{d+2} X \sqrt{-g}  \nabla^\mu \t J_\mu^S  \bigg] \\
&= \exp \bigg[ - \frac{i}{e^2}\int_{\S^+ - \S^-} \dt\S^\nu  \nabla^\mu \t  F_{\mu\nu}  \bigg] ,
\end{split} 
\end{align}
where we integrated by parts and used the fact that $F$ vanishes on $i^0$ to obtain the second line. Combining with the matter dressing Eq.~\eqref{wilson-hard-1} and using Maxwell's equation Eq.~\eqref{Maxwell_Eq} only on $\S^\pm$ and not on $i^0$ (where the matter current is already assumed to vanish) to ensure the hard modes associated with the matter current are on-shell but the soft modes are off-shell \footnote{Because the soft effective action to which we are matching the Wilson line dressing is an off-shell quantity involving the soft modes, we do not impose equations of motion for the soft modes on the celestial sphere at $\CI^\pm_\mp$, or equivalently $i^0$. Indeed, if we impose equations of motion everywhere, we get $\CW^\tree_S \CW^\tree_M = 1$, which implies the matching condition $Q_\t^+ = Q_\t^-$ (see Eq.~\eqref{wilson-soft}). This is precisely the result of \cite{Campiglia:2017mua}. We thank Adam Ball for pointing this out.}, we obtain
\begin{equation}
\label{wilson-soft}
\begin{split}
\CW^\text{tree}_M  \CW^\text{tree}_S 
&= \exp \bigg[ - \frac{i}{2e^2} \oint_{\CI_-^++\CI_+^-} \!\! \dt S^{\mu\nu} \, \t F_{\mu\nu} \bigg]   \\
& = \exp \bigg[ - \frac{2i}{e^2} \oint_{\CI^+_- + \CI^-_+} \!\! \dt^d x \, |r|^d \t F_{ur} \bigg] . 
\end{split}
\end{equation}
Comparing this with Eq.~\eqref{charge}, we find
\begin{equation}
\begin{split}
\label{wilsondressing_softcharge}
\CW^\text{tree}_S  \CW^\text{tree}_M = \exp \Big[ i \big( Q^+_{\t} - Q^-_{\t} \big) \Big] .
\end{split}
\end{equation}
Finally, using Eq.~\eqref{equivalence_1}, we immediately prove Eq.~\eqref{equivalence_3}.

We now present the second interpretation of the Wilson line dressing given in Eq.~\eqref{equivalence_3}, which relies on only considering Wilson lines given in Eq.~\eqref{soft_dressing} to be localized on the celestial sphere. Letting $\bar\gamma_k$ be such Wilson lines, the analogue of Eq.~\eqref{soft_dressing} is
\begin{align}
\begin{split}
    U_k &= \exp \bigg[ i Q_k \int_{\bar\g_k} \dt x^a A_a(x) \bigg] .
\end{split}
\end{align}
Utilizing the decomposition Eq.~\eqref{gaugefield_decomp} again, the $\hat A_a$ term in the soft limit does not contribute at tree-level, so it suffices to replace $A_a$ with $\p_a \t$, yielding
\begin{equation}
\begin{split}
    U_k \big|_\text{soft} &= \exp \bigg[ i Q_k \int_{\bar\g_k} \dt x^a \, \p_a \t(x) \bigg] \\
&= \exp \Big[ i Q_k \big( \t(x_k) - \t(x_0) \big) \Big]  ,
\end{split}
\end{equation}
where $x_0$ and $x_k$ are the endpoints of the Wilson line $\bar\g_k$. If all the particles in the scattering amplitude are massless, then the corresponding Wilson line dressing is simply
\begin{equation}
\begin{split}
\label{massless_dressing_1}
    \CW^\tree_{M} = U_1 \cdots U_n \big|_{\text{soft}} = \exp \left[ i \sum_{k=1}^n Q_k \t(x_k) \right] ,
\end{split}
\end{equation}
where the dependence on the base point $x_0$ has dropped out completely due to charge conservation. This means the dressing factor $\CW_M^\tree$ does not depend on the initial point from which the Wilson lines originate. Indeed, the celestial Wilson lines ${\bar \g}_k$ can therefore be related to the bulk Wilson lines $\g_k$ by moving the initial point of $\g_k$ (namely, the origin) to $x_0$ on the celestial sphere.

The generalization of Eq.~\eqref{massless_dressing_1} to the massive \emph{and} soft Wilson line dressing is rather straightforward. Unlike massless particles, which pierce the celestial sphere at a single point, massive particles and soft photons have a non-local charge distribution. The corresponding distributional Wilson line is
\begin{equation}
\begin{split}
\label{dressing_chargedensity}
    \exp \bigg[ i \oint_{\mss^d} \!\! \dt^d x \, \rho(x) \t(x) \bigg] ,
\end{split}
\end{equation}
where $\rho(x)$ is the charge density on the celestial sphere associated to massive particles and soft photons. For massless particles, the charge density is $\rho_k^\text{massless}(x) = Q_k \d^{(d)}(x-x_k)$, which reproduces Eq.~\eqref{massless_dressing_1}. For massive particles, it is given by $\rho_k^\text{massive}(x) = Q_k \CK_d(m_k/\o_k , x_k ; x )$, where $\CK_\D$ is the bulk-to-boundary propagator in $\ads_{d+1}$ \cite{Kapec:2021eug}. Further including the charge density for the soft photons, the total Wilson line distribution becomes
\begin{align}\label{soft-dressing-final}
\CW_S^\tree \CW_M^\tree = \exp\left[ i \oint_{\mathbb S^d} \!\! \dt^dx \,\rho_{\text{total}}(x)  \t(x) \right] ,
\end{align}
where $\rho_{\text{total}}(x)$ is the total charge density. The argument of the exponential is the integral of the charge density-weighted by $\t(x)$, which, by our charge convention, is precisely $ i (Q_\t^+ - Q_\t^-)$. Hence, we see that Eq.~\eqref{wilsondressing_softcharge} holds once more.

\section{Discussion}\label{sec:discussion}

The primary objective of this paper is to show that there are four different objects in the IR sector of abelian gauge theories that are identical, given that the large gauge charge $Q_\ve$ is parametrized by the Goldstone mode so that $\ve = \t$, and that soft photons are dressed. While our analysis is purely in the context of abelian gauge theories and at tree-level, we suspect due to the universality of IR dynamics that these equivalences extend to nonabelian gauge theories and, most importantly, gravity. Furthermore, \cite{He:2024ddb} proved that the equivalence between the edge mode action and soft effective action survives loop corrections, and it would be worthwhile to explore if the other sides of the IR diamond survive loop corrections as well.

In our analysis, there are two important novel features we emphasize, which will be especially interesting in the context of gravity.  The first is the promotion of the large gauge charge parameter $\varepsilon$ to the dynamical Goldstone mode.  As a result, the large gauge charge involves a mixing between the (shadow transform of the) Goldstone $\widetilde C^a(x)$ and the soft photon $N^a(x)$. The large gauge charge is now a composite operator, and in general its vacuum expectation value will be non-vanishing.  In gravity, the analogous charge is the supertranslation charge, and it would be interesting to compute its expectation value and variance to verify the results of \cite{Verlinde:2019ade,Banks:2021jwj,Verlinde:2022hhs} obtained using other means. Related ideas were explored in \cite{Kapec:2016aqd}. 

The second novel feature above is the introduction of a Wilson line distributional dressing for the soft photon. Such a dressing is quite natural for dressing massive charged particles, whose charge distributions are not delta function localized on the celestial sphere. Therefore, because soft photons similarly have a nontrivial charge distribution on the celestial sphere, we expect them to also have a distributional Wilson line dressing. It is not immediately obvious how to interpret such a dressing from a conventional field theoretic perspective, and we leave a better understanding of this for future work.

Ultimately, we would like to extend our above analysis to nonabelian gauge theories and (perturbative) gravity. Unlike soft photons, soft gluons and (at least at subleading order) soft gravitons carry nontrivial global charge in addition to local charge, and it would be interesting to explore the consequences of this difference. Ironically, because the leading soft behavior of abelian gauge theories and gravity is similar, we expect the case with gravity to be possibly simpler. For instance, the relationship between the soft effective action and the supertranslation charge is nearly identical to our analysis above \footnote{The connection between the supertranslation charge and the ``modular Hamiltonian'' studied in \cite{Verlinde:2022hhs} is much less obvious and is the focus of \cite{He:2024vlp}.}. On the other hand, for nonabelian gauge theories, the soft effective action is not even known, and has only been conjectured in \cite{Kapec:2021eug, Magnea:2021fvy, Agarwal:2021ais}. We hope to pursue a satisfying understanding of the four corners of the diamond for these cases.

\section*{Acknowledgements}

We would like to thank Laurent Freidel and Ana-Maria Raclariu for productive conversations, and Adam Ball for useful feedback regarding our preprint. We would especially like to thank Allic Sivaramakrishnan for collaborating in the early stages of this work. T.H. and K.Z. are supported by the Heising-Simons Foundation ``Observational Signatures of Quantum Gravity'' collaboration grant 2021-2817, the
U.S. Department of Energy, Office of Science, Office of High Energy Physics, under Award
No. DE-SC0011632, and the Walter Burke Institute for Theoretical Physics. P.M. is supported
by the European Research Council (ERC) under the European Union's Horizon 2020 research and
innovation programme (grant agreement No 852386). The work of K.Z. is also supported by a Simons Investigator award.

\bibliography{YMbib}
\bibliographystyle{apsrev4-1}

\end{document}